Chapter 6

# MICROARRAY DATA MANAGEMENT

*An Enterprise Information Approach: Implementations and Challenges.*

WILLY A. VALDIVIA-GRANDA[1*]; CHRISTOPHER DWAN[2]
[1]Orion Integrated Biosciences Inc. New York, USA

[2]BioTeam Inc. Massachusetts, USA

**Abstract:**

The extraction of information form high-throughput experiments is a key aspect of modern biology. Early in the development of microarray technology, researchers recognized that the size of the datasets and the limitations of both computational and visualization techniques restricted their ability to find the biological meaning hidden in the data. In addition, most researchers wanted to make their datasets accessible to others. This resulted in the development of new and advanced data storage, analysis, and visualization tools enabling the cross-platform validation of the experiments and the identification of previously undetected patterns. In order to reap the benefits of this microarray data, researchers have needed to implement database management systems providing integration of different experiments and data types. Moreover, it was necessary to standardize the basic data structure and experimental techniques for the standardization of microarray platforms. In this chapter, we introduce the reader to the major concepts related to the use of controlled vocabularies (ontologies), the definition of Minimum Information About a Microarray Experiment (MIAME) and provide an overview of different microarray data management strategies in use today. We summarize the main characteristics of microarray data storage and sharing strategies including warehouses, datamarts, and federations. The fundamental challenges involved in the distribution, and retrieval of microarray data are presented, along with an overview of some emerging technologies.

# 1. INTRODUCTION

A microarray is a high-density two-dimensional matrix where thousands of nucleic acid, proteins or tissues are immobilized on the surface of a glass slide, nylon filter, or silicon wafer. The primary purpose of a microarray is to perform biological screening experiments at the whole genome scale. Each 'spot' represents a single biochemical assay 'probe' against a particular object of biological interest, perhaps measuring the expression level of a gene, or the binding efficiency of a genomic regulatory element. Using this technology, researchers effectively perform tens of thousands of measurements in parallel.

There are many ways to perform the "spotting" process by which samples are placed on a microarray. In contact printing, mechanical pins can be used to robotically transfer micrograms of probe from storage trays onto slides or membranes. In non-contact printing, ink-jet style printing techniques spray various amounts and configurations of probe. Finally, *in situ* synthesis using photolithographic methods can build cDNA or RNA strands, residue by residue. Because of the distinction between sample spotting and photolithography, the latter are sometimes referred to as "DNA Chips." For the purposes of this document, we refer to both techniques as "microarrays." Both contact and non-contact printing give spots of 100 µm in diameter, while photolithography spots are about 20 µm. These processes produce microarrays with spot densities from 10,000 to 250,000 spots per $cm^2$.

Because the spots printed on an array surface are typically less than 200 µm in diameter, microarrays need to be read by specialized scanners. Most commercially available microarray scanners are inverted florescent microscopes that acquire data at two wavelengths (generally used to record a test and a control signal) using 532 nm (17 mW) and 635 nm (10 mW) lasers. The output of this process will be an image file (~5 Mb) and a text file (~1.5 mb). The text file provides primary data on the intensity ratios of the two wavelengths, averaged over the area of each spot. In order to assess the contribution of experimental noise and error inherent in this new technology, it has become standard process, in contact and non-contact array manufacture, to place an abundance of replicates of each probe on a single microarray. In addition, most experiments involve multiple copies/instances of each microarray. A single microarray experiment might involve measuring the expression of a particular set of genes at one hour intervals during the 24 hour following exposure to some environmental stress. This would produce, with even modest experimental redundancy, nearly half a gigabyte of primary data.

In less than a decade, microarrays have become a widespread technology used for the exploration of molecular activity of biological systems. Since their development, more than 12,000 publications have relied on them for primary experimental results. This demonstrates their impact on biological sciences. The wide use of microarrays is the result of two factors: the decreasing cost of reagents and instruments, and the fact that they are so effective as an experimental technique. Today, the end cost to a researcher, to measure the expression of a gene is approximately $0.05 [1]. Of course, this assumes that the researcher is willing to measure gene expression in batches of tens of thousands. The high number of probes permits the exploration of complete genomes, including non-coding regions [2, 3]. The diversification of microarray technology to include tissues [4-6], proteins, and peptides permits interrogation of the molecular activity of the cell at many levels of resolution [1, 7].

An increasing number of laboratories are using microarray-based analysis for disease fingerprinting, toxicological assessment, single nucleotide polymorphism (SNP) analysis, reconstruction of signal transduction pathways, and phylogenomic and epigenetic analysis [8-12]. Microarrays are also ideal for fast, sensitive, specific, and parallelized detection and diagnosis of microorganisms [13]. This has applications in primary research, clinical medicine, and biodefense. Several researchers have used microarrays for the genotyping of influenza viruses [14, 15]; drug resistant HIV-1 [16]; polioviruses [16]; human papiloma [17]; RNA respiratory viruses [18, 19]; hepatitis B and C [20]; and African swine fever [17, 21].

In the early stages of microarray technology development, researchers recognized that, due to the size of the datasets involved, computational analysis would be required to properly exploit the information. Early microarrays were very expensive, and for this reason several researchers restricted themselves to analyzing datasets published by others. At this stage, the sharing of microarray data was mainly accomplished by exchanging flat files. This enabled progress, despite the lack of standards to exchange genomic information. The key to this success, however, was the personal communication between the researcher who had done the physical experiment and the one doing the analysis. The use of flat files coupled with the lack of direct communication has several limitations. The primary problem is in the exchange of experimental parameters, the “metadata” without which the raw data is meaningless. Most microarray experiments are composed of many different gene expression data files. To understand the biological significance of its content it is necessary to integrate several types of genomic information (e.g. the assignment of the molecular function of genes, the history of the samples used on the microarray, the batch and lot numbers of the slide, the settings of the scanner, and so on). There is also difficulty involved in retrieving a subset of genes and expression values from flat files without extensive script programming information. Nonetheless, a main advantage of the use of flat file format is that microarray data could be provided *as is*.

Spreadsheets are another file format used to store and share microarray data. This format not only allows sorting and filtering, but also makes it possible to perform basic calculations and to produce graphical representations using add-ins and collections of macros developed specifically to analyze microarray data [22-24]. Unfortunately, spreadsheets are difficult to update or manage remotely. Moreover, the proprietary format of this platform has limited impact in the extensive exchange of microarray data. For this reason, researchers typically link spreadsheets with web pages in the context of their publication. While requiring little effort to implement, the content and quality of the information contained within the spreadsheets is dependent on the algorithms used for normalizing, filtering and analyzing. Of course, the above mentioned limitations of metadata transfer apply just as much to spreadsheets.

The wide availability of microarray data has fueled the development of *exploratory research* and the generation of new hypothesis about specific biological processes based on the analysis of large amounts of data. A typical example is the dataset published by Golub et al. [25]. It has been analyzed by different researchers using a variety of statistical and computational methods [26-34]. Because different algorithms applied to the same data can provide new insights about a particular biological process, the integration of different experiments through automated database management systems can have a significant impact on understanding/interpretation. This phenomenon has already been seen with databases storing genomic and protein sequence data. With the emergence of the study of biological systems in a holistic manner (also known as biocomplexity or systems biology), the analysis of microarray data is placed in conjunction with that of the other *omic* datasets [18, 35-37]. This has enabled the development of multi-resolution molecular maps of specific biological processes.

Currently, around 3% of more than 400 biological databases store microarray data [35, 38, 39]. However, many researchers performing microarray experiments are unfamiliar with database concepts and perceive data management systems as *black boxes* for data input and retrieval. With this in mind, the objective of this chapter is to introduce the reader to the basic concepts related to the storage, use, and exchange of microarray data including:

- A description of the use of ontologies to provide a structured vocabulary for cataloging molecular components of the cell as well as details about microarray experiments.

- An overview of different data models to exchange genomic information, including the minimum information about a microarray experiment (MIAME) standard.

- A description of different microarray database management systems. and the main characteristics of microarray data integration projects, including data warehouses, datamarts, and federated databases.

- An overview of new developments in data storage, exchange and high performance computing for the implementation of enterprise data and microarray knowledge management systems.

- A highlight of the main challenges and opportunities related to the development of new exchange systems and the access to data streams.

## 2. MICROARRAY DATA STANDARIZATION

The issue of data standards, integration, and interoperability has long been of interest to biologists. DNA and protein sequence formats like those used by Genbank, Swissprot, and PDB reflect such need. The structure of this information allows researchers to write specific parsers to retrieve subsets of information which are in an XML or flat file format. When analyzing nucleic or amino acid sequences, researchers are interested in obtain information other than the sequence data. For example, they might want to know about genomic context: the length of the open reading frame, the frequency and location of known introns, the chromosomal location, and any putative molecular function. In most cases, this information is stored in separate databases.

Because most microarray experiments measure the transcriptional activity of genes, the information about a particular gene is very relevant. Additionally, since the variability and reliability of the experiment is affected by multiple factors, microarray analyses require detailed information about the experiment itself before the raw data can be interpreted at all.

A typical experiment using microarrays involves a team of researchers. Each member has skills in a particular process: from tissue preparation, microarray production (or selection from a corporate provider), RNA extraction and cDNA dye labeling, operation of the microarray scanner, normalization and data analysis. Some of these steps may even be outsourced to external sources. During each step, different sources of noise and variability are introduced. As a result, missing data, outliers and variability across replications and laboratories is very common. A researcher integrating different microarray datasets must know the strengths and weaknesses of each, as well as their relative level of appropriateness for the current investigation.

To integrate different databases, we must establish points of reference in the metadata and compare the data from various experiments in light of those reference points. Comparing free text definitions is very difficult. Different research groups may come up with different definitions for a particular experiment or biological process. They also may use very similar words to describe fundamentally different processes. For instance, a researcher might use the term DAG to mean directed acyclic graph, but for most cell biologists it will be the shorthand for diacylglycerol, a key intracellular signaling component in the calcium transduction cascade. Therefore, when integrating genomic information, the reader should be very aware that biology is a massive and dynamic field of experimental study. Word meanings are not stable between experimental domains, and as new discoveries are made, new data definitions of genes, genomes and biological systems emerge.

To facilitate the retrieval of genomic information and the exchange of microarray data, researchers recently have begun to agree on a common set of terminologies and a minimum set of parameters that should be used to describe experiments involving this technology. The formation of government initiatives for the standardization of protocols and reagents, as well as the use of microarrays in clinical studies and for the diagnosis of pathogens, has prompted this need. In order to provide the reader with the overall understanding of the significance of these implementations, we will review concepts related to the gene and microarray ontologies and the Minimum Information About a Microarray Experiment (MIAME) standard.

## 2.1 Gene Ontologies

The abstraction of real-world concepts is very important in the creation of information exchange systems and the management of knowledge. Most applied mathematics is based on this fundamental truth. In the early 1990's the artificial intelligence community developed a framework for the use of controlled vocabularies to capture and formalize the knowledge in a particular domain. Ontologies specify the terms or concepts and relationships among terms and their intended correspondence to objects and entities that exist in the world. Domain ontologies are specialized collections of names for concepts and relations organized in a particular order. These descriptions and rules are accepted by a community in an interdependent fashion. They allow computer generated queries to filter and retrieve information based on user defined constraints [40-42].

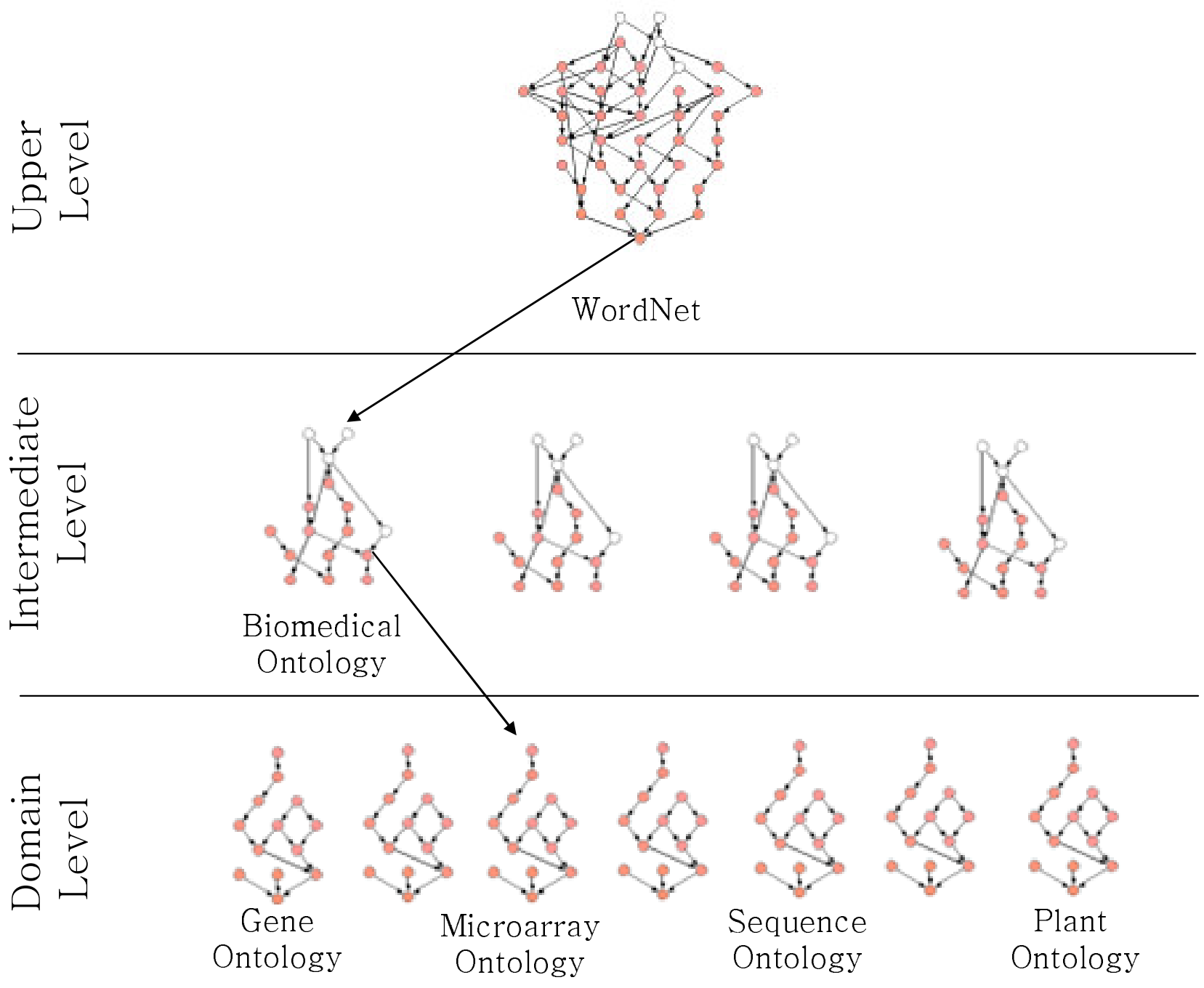

*Figure 1.* Abstraction of different ontology levels. Adapted from Soldatova and King [46]. While upper categories are useful for generating the structural backbone of the intermediate and domain ontologies, the domain hierarchy requires more complex logical expressions.

The implementation of ontologies can be accomplished using specialized development environments, including the Knowledge Interchange Format (KIF), Ontolingua, WebOnto, $\mu$Kosmos, Cyc and Protégée. However, ontologies vary in their coverage, quality and resolution. From an implementation point of view, three main types of knowledge representation can be can be distinguished:

- **Upper ontologies:** also called high-level, core or reference ontologies, describe common general concepts across different communities (*e.g. SUMO*, and WorldNet).

- **Intermediate ontologies:** are shared ontologies among domains that allow for scalability and join domain and upper ontologies.

- **Domain ontologies:** are restricted in their scope and coverage to interest of a particular domain (*e.g.*: plant ontology, human anatomy ontology, gene ontology, microarray ontology). Domain ontologies join and leave intermediate and upper ontologies and are in constant development.

The sequencing of genes and genomes led to the proliferation of many biological databases. The information contained in these repositories was designed to be populated and accessed by humans, rather than by computers, and was littered with inconsistencies. The functional role of genes tended to be annotated as free text phrases. Many of these where classified into arbitrary categories. At the very least, competing spellings of common terms made simple text searching unwieldy. As a result, it was difficult to search the databases for the function of a particular gene or biological process. Integrating these repositories was a Herculean task, usually only undertaken within a fairly small community surrounding a particular area of research.

To address these issues, Schulze-Kremer [40] proposed the use of ontologies to provide a standardized description of objects and process related to molecular biology. An ontology for the molecular function, biological process and cellular components of genes was proposed by The Gene Ontology Consortium (GOC) [43]. Their effort lead to the implementation of independent terminologies for species, as well as classifications related to genes.

The GO now has approximately 17,000 terms and several million annotated instances describing how gene products behave in a cellular context. A particular term is linked directly to some datum in a public database. The GO is used by at least 30 major bioinformatic databases serving researchers interested in more than 140 organisms. Each term in the gene ontology is accessible by a unique identifier (GO ID) and every annotation must be attributed to a source which may be a literature reference or a computer generated annotation.

In a relatively short time, the GO has been adopted by the biological community. Its impact is due to the strictness and expressiveness that allows software architectures to compute and associate biological information from disparate databases. The GO has also gained considerable credibility for simply starting with a large, overlapping set of definitions, rather than haggling over an exact data modeling standard. For these reasons, the GO has become the *de facto* standard for biological database ontologies[44].

The graphical representation of the gene ontology is made as a semantic net or conceptual graph (both of which are instances of a directed acyclic graph- DAG). A DAG consists of a set of nodes, and a set of edges. An edge is a pair of nodes and the order of the nodes in the edge makes a difference; i.e. the edge (a,b) is different from the edge (b,a). This type of representation is ideal for critical path analysis and for understanding the relationships between different hierarchical categories of the gene ontology (Figure 2).

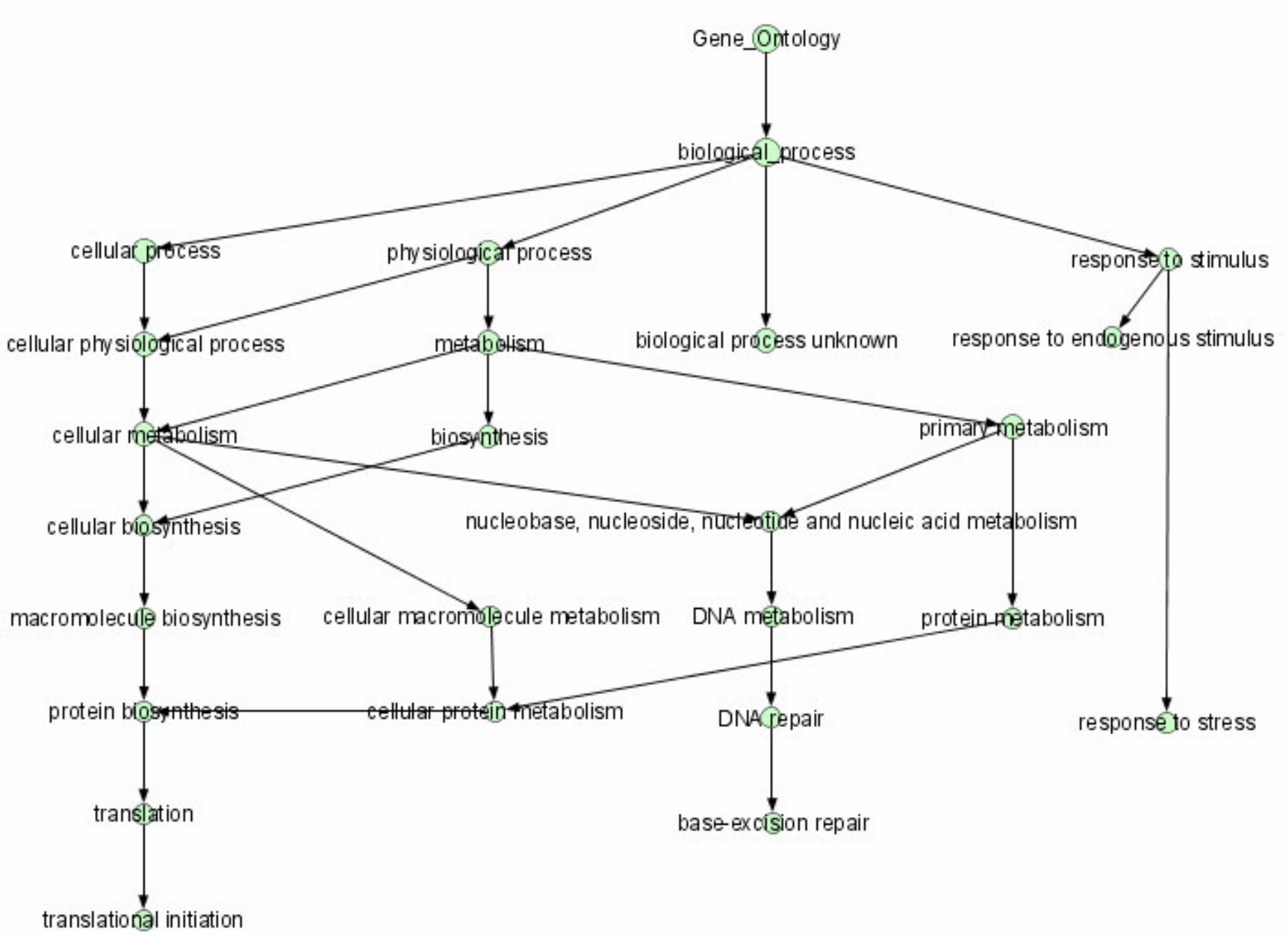

*Figure 2.* Figure 2. DAG representation of the ontological arrangement of the biological function of twenty genes of malaria (*Plasmodim falciparum*). Please note that several nodes are shared by different parts of the tree.

## 2.2 Microarray Ontologies (MO)

An effective microarray database should allow researchers involved in data analysis to pose a query in terms used by an experimentalist, and retrieve a unified dataset from multiple sources. This, however, requires knowledge of the experimental parameters affecting the reliability and the quality of a particular microarray experiment. To properly join the concepts and definitions describing these experiments and to facilitate automated querying and exchange of this microarray data, a group of public and private researchers formed the MGED Ontology Working Group [45]. This effort is standardizing the terminology required to publish a microarray experiment. The MGED Ontology Working Group is composed of computer scientists, developmental biologists, toxicologists, and the whole microarray community. This group is collaborating on the makeup of a microarray ontology (MO) using each member's knowledge for their area of expertise. MO uses the Protégée development environment and is divided into two parts [46]. The *core layer* is a static ontology describing only essential concepts about microarray experiments. This layer is intended to be relatively static. The *extended layer* describes concepts related to microarray experiments and changes as biological knowledge and microarray platforms evolve (Figure 3).

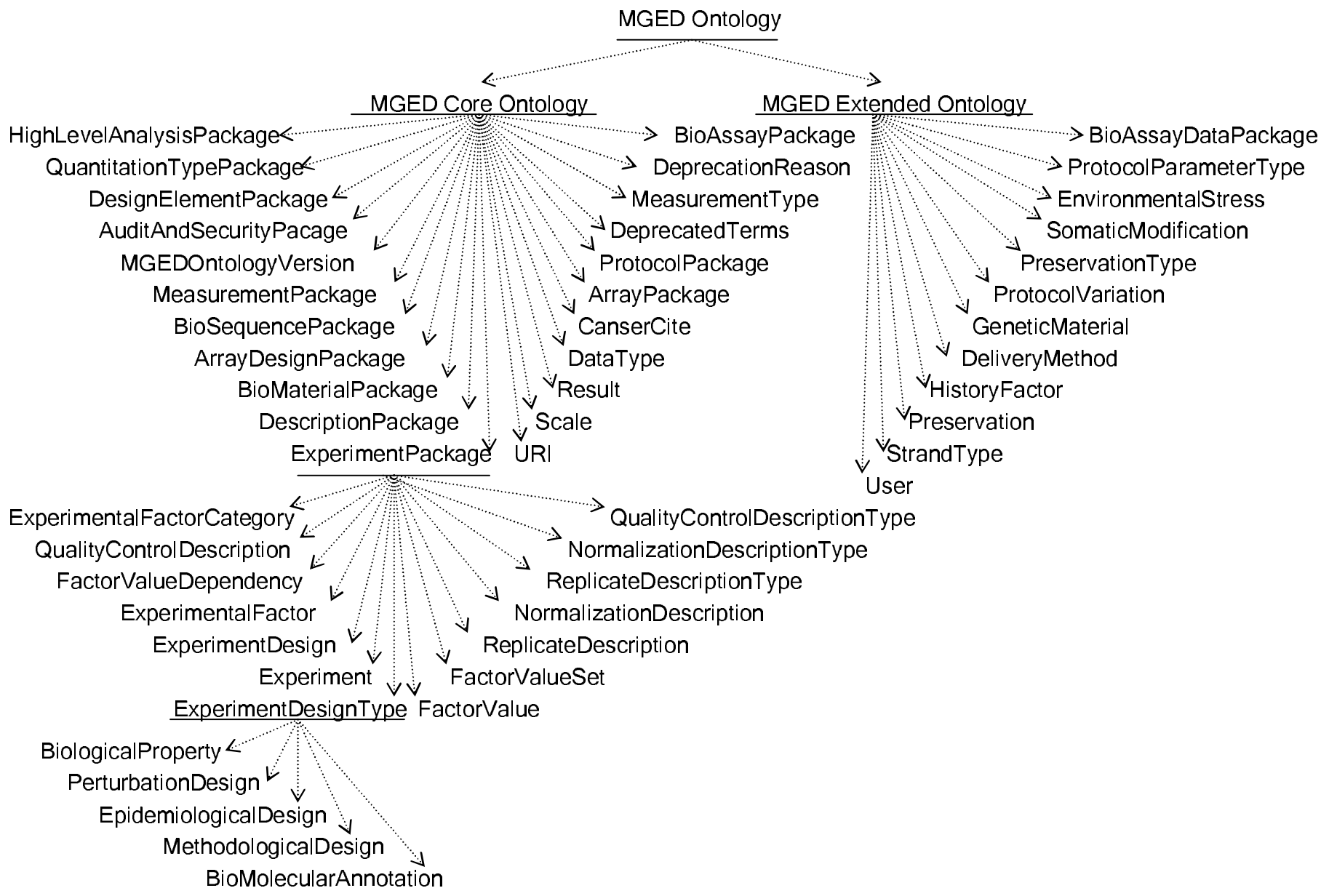

*Figure 3.* The figure shows the detailed concept hierarchy for <ExperimentPackage> and <ExperimentDesignType>. Courtesy of L. Soldatova and R. King [46].

MO and GO are the first attempts to formalize in a consistent way the description of experiments and the molecular components of the cell. Although the design and implementation of these integration infrastructures is still under development; Soldatova and King [46] have pointed several awkward linguistic issues in the naming policy and the design of the GO and in particular, the MO. The fact that GO and MO do not contain enough terms to describe actual microarray experiments or biological processes limits its mapping, alignment and merging to intermediate and upper ontologies. Also in several instances MO uses the same name at different levels of abstraction and allows multiple inheritances of properties. Despite the obvious limitations, MO and GO avoids subjective interpretations of the meaning of microarray experiments and gene descriptions. However, as new experiments become available, its redesign or reconstruction is becoming obvious.

## 2.3 Minimum Information About a Microarray Experiment (MIAME)

To achieve the integration of microarray datasets, researchers need to agree not only on the GO (what we are using or observing) and MO (what data we are collecting), but also on the manner in which the experiment is being conducted. There is a considerable variability in both reagents and reference controls, and therefore, it is difficult to compare microarray data generated by different laboratories [7, 47]. The MIAME strictly defines each of the parameters that should be reported in order to provide sufficient information to allow an outsider to interpret the experiment [48]. Most importantly, the MIAME is facilitating microarray applications in clinical and diagnostic settings. The MIAME- annotation has six major sections:

- Experimental design
- Array design
- Samples
- Hybridization
- Measurements and
- Normalization

An updated summary of the MIAME guidelines is available in the MGED society website. In addition, the MIAME is also serving as a blueprint for standardization specific type of experiments [49, 50]. MIAME-Tox includes descriptors for the inclusion cell types, anatomy terms, histopathology, toxicology, and chemical compound nomenclature in the context of toxicogenomics and pharmacogenomics research [51-53].

## 3. DATABASE MANAGEMENT SYSTEMS

The storage, exploration, and exchange of microarray data require computer systems capable of handling many simultaneous users, performing millions of data transactions, and transfering many terabytes of data in a secure and reliable way. Fortunately, there is a robust field of software development known as database management systems (DBMS) dedicated to exactly this task. DBMS tools are frequently referred to as "databases," which leads to confusion between the software infrastructures used to manage the data (the DBMS) and the collection of data being managed. In this section, we are discussing DBMS software. Examples include products such as Oracle, Sybase, DB2, and MySQL. The use of a DBMS can provide many benefits: secure access to both journal published and unpublished data, the elimination of redundant, inconsistent and outdated information, reliable data storage and retrieval, data provenance and historical recovery.

There is no reason to limit a DBMS to storing only primary data. It is also possible to use DBMS to store "data about data," or metadata. However, metadata requirements must be identified a priori, and should include scientific, computing and administrative considerations. Using the metadata, researchers can compose queries that incorporate the quality, condition, or even physical location. From an implementation point of view, we can divide metadata into:

- **Technical metadata:** This information is primarily used to support the work of the staff that is deploying and implementing a particular DBMS. Technical metadata describes the physical organization of the database, the access policies, user accounts, and the integrity constraints that allow the system to operate effectively.

- **Microarray metadata:** In the context of this document, is the data annotated using the MIAME and GO standards, including the use of the MO.

Using a DBMS, one can vastly accelerate the process of data exchange and analysis and therefore, researchers can improve their understanding of specific biological processes. However, in contrast to data shared via flat files, data stored in a DBMS must conform to specific rules within a mathematical framework known as the *data model.*

A *data model* is a conceptual representation of the mathematical rules that define the relationships between different components of a database. In other words, the data model defines what data is required, and how it should be organized. Over the years, database researchers have proposed six main data models: file processing, hierarchical, network, relational, object-oriented, and the object-relational. In this document, we focus in on the last three data models which are commonly used to exchange microarray information.

## 3.1 Relational Data Model (R-DM)

The relational data model was developed by Codd (1970). The main idea behind this approach is the representation of data in two dimensional tables. This data structure in many ways drove the enterprise adoption of computers in financial, business and research applications. The basic elements of the relational data model are the table (or *relation*) that is composed of rows (*tuples*) and columns (*attributes*). Each table has a unique *attribute* known as the *primary key* that identifies a *tuple*. Relationships between two tables are made by matching their primary key values. While, the primary key of each table can never be a null value, a *foreign key* permits the association of multiple tables defined by a *schema*. The term schema is often used to refer to a graphical depiction of the database structure (Figures 4, 6 and 7) and defines the fields in each table, and the relationships between fields.

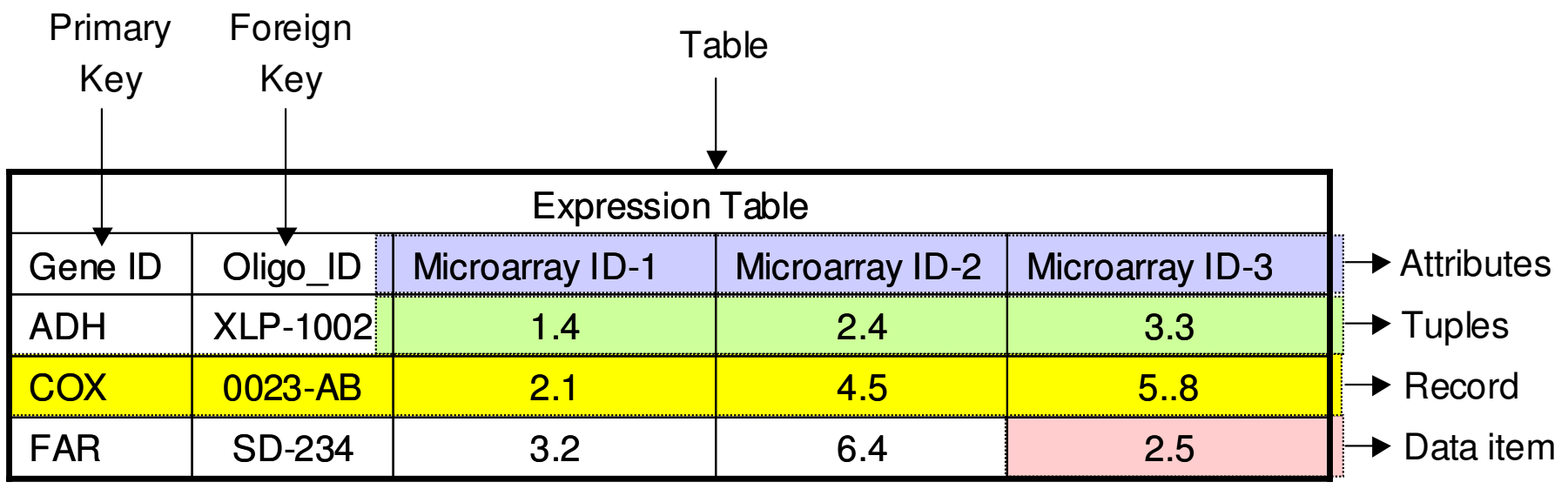

*Figure 4.* Schematic representation of different elements of the relation or table.

The mathematical basis of the relational model results in a series of algebraic operations to manipulate the data. These include SELECT, PROJECT, JOIN, PRODUCT, UNION, INTERSECT, DIFFERENCE and DIVIDE. These commands allow the user to combine, filter and retrieve the attributes of different tables. For example, to retrieve the expression values of a cluster of 300 genes from only 12 of the 48 time point series experiment, the user needs to select the 300 primary keys that identify those genes, and join those keys with attributes from each of the 12 tables. The basic query specification consists of "SELECT attributes (300 genes IDs) FROM table names (time points 1 to 12)". The most common language for expressing queries of this sort is the Structured Query Language (SQL). However, it is necessary to clarify that there are several variants of the relational model. There are several software implementations capable of performing queries using the relational model. These relational database management systems (RDBMS) include: Oracle, Sybase, DB2, Informix, PostgreSQL, MySQL, and MS-Access.

### 3.1.1 Notation of the Relational Model

Whether the implementation of a relational database is intended to serve the needs of a large number of researchers or small workgroup, the planning of its design is an important step ensuring future performance of the database. Notation is a logical and graphical design technique often used to allow designers, implementers and users to understand in advance the mathematical relationships encoded by the database. The notation is also a valuable graphical representation that facilitates the redesign and update of the database.

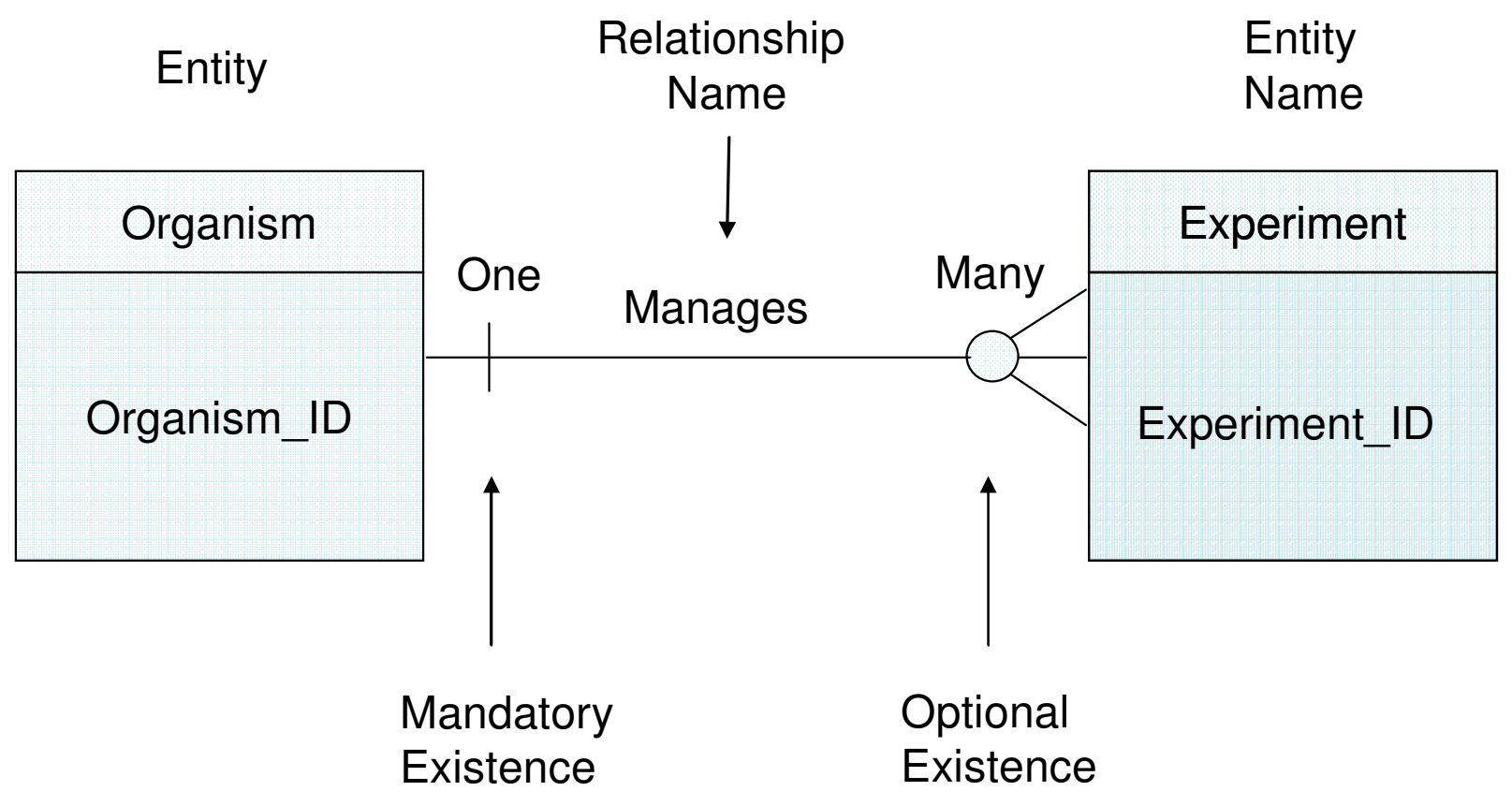

*Figure 5.* Basic components in the notation of the relational model. The connectivity of a relationship describes the mapping of associated entity instances in the relationship. The cardinality of a relationship is the actual number of related occurrences for each of the two entities. The basic types of connectivity for relations are: one-to-one, one-to-many, and many-to-many.

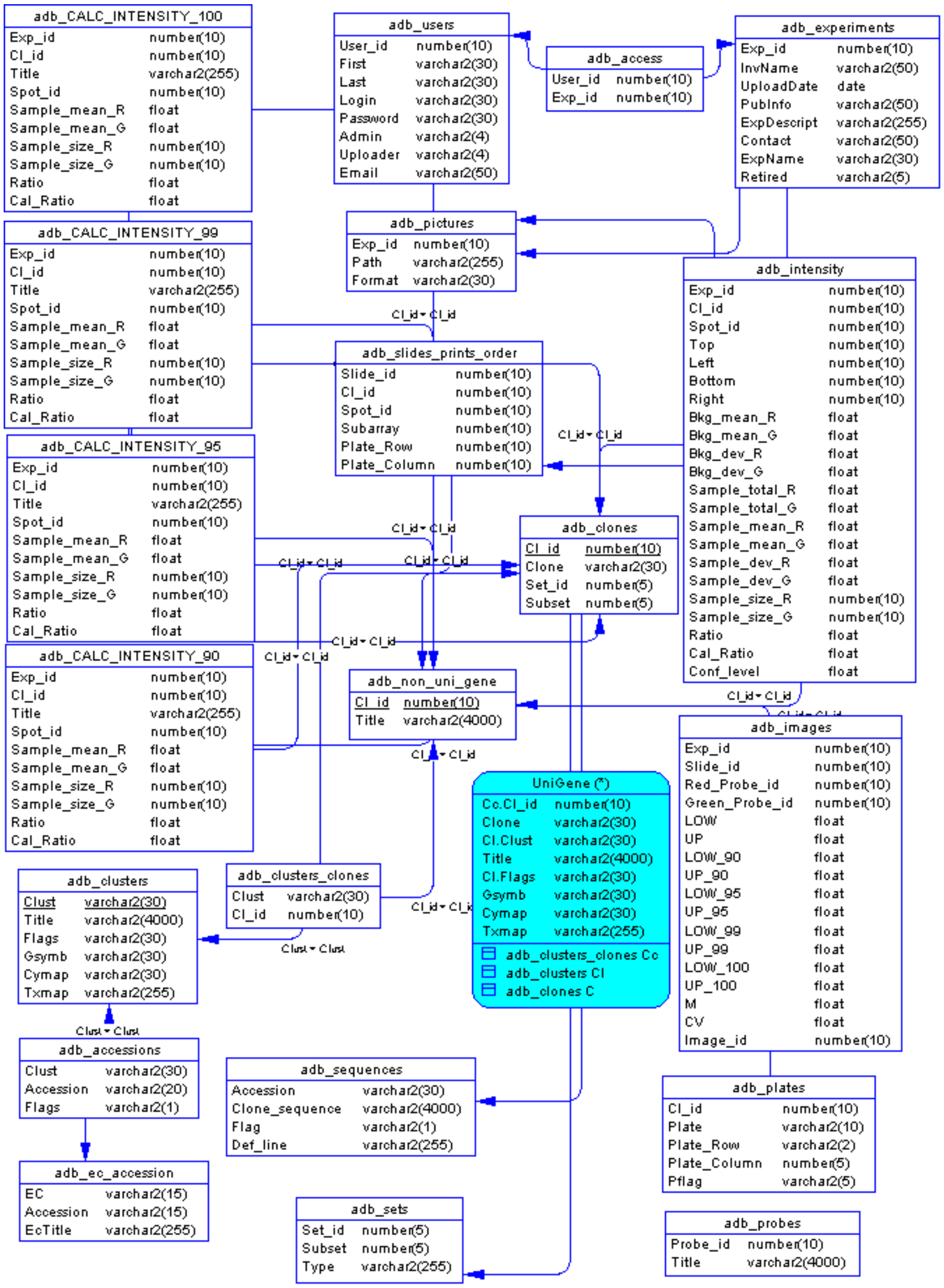

*Figure 6.* Diagram of a different tables and the overall relational microarray database structure of the ArrayDB [50]. Courtesy of A Baxevanis. The schema is partitioned into non-disjointed sub-schemas according to the needs of the different divisions within the enterprise and the different sub-applications. Relationships between abstract categories are shown by arrows between the categories' rectangles. The name and type of the relation are indicated on the arrow.

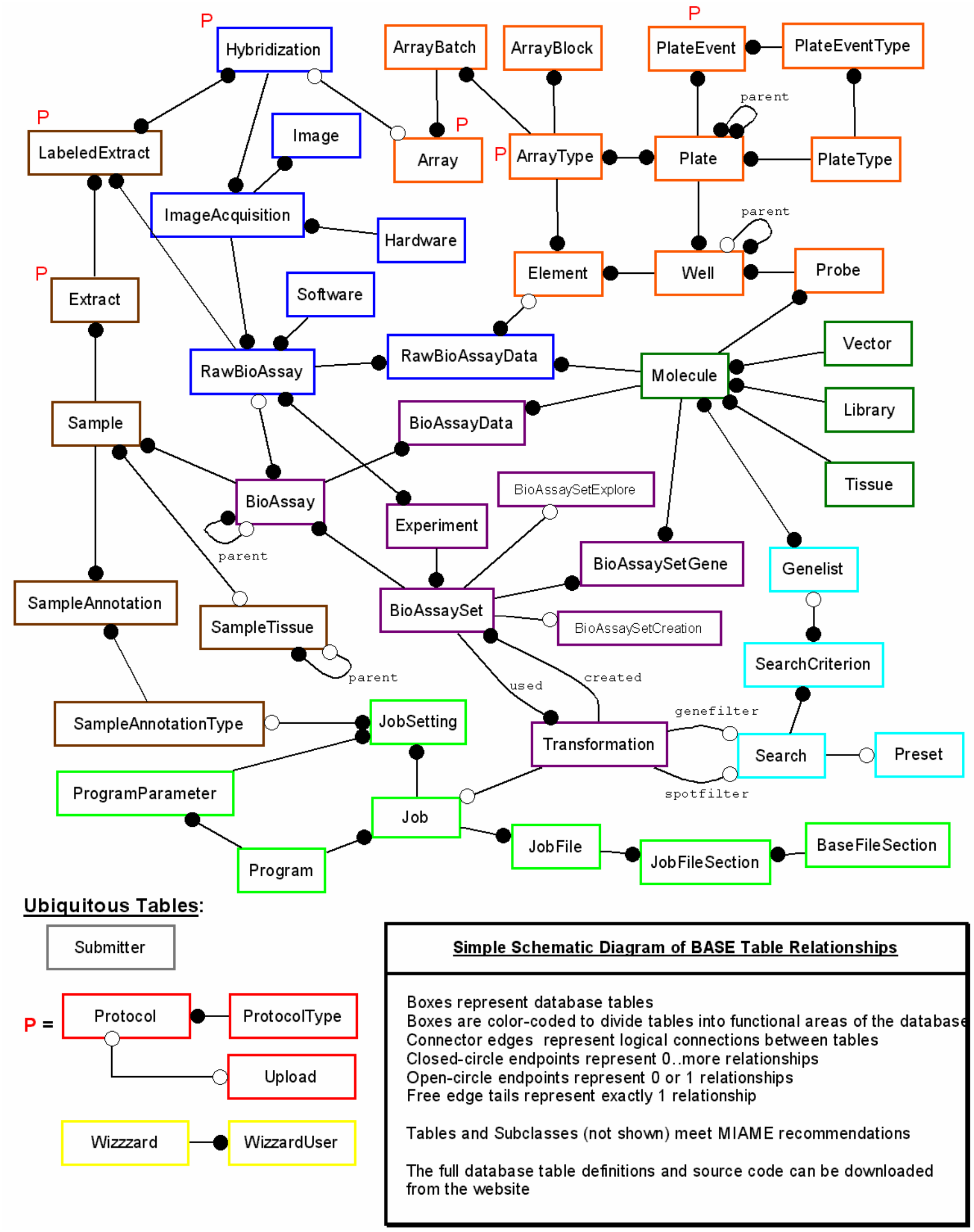

*Figure 7.* Schema notation of the BioArray Software Environment (BASE) [54]. Courtesy of L Saal. All BASE source code is publicly available to academic and commercial sites under the GNU General Public License. Notice that the design of BASE involves first creating a schema to describe the structure of the database and the rules for the interactions for each table.

### 3.1.2 Limitations of the Relational Model

The relational model is simple to understand and use, even for those who are not experienced programmers. However, the use of the R-DM is poorly suited to the integration of microarray experiments with other types of genomic information. The relational model does not handle certain forms of data well including images, sequence data and digital documents. These limitations can restrict the scalability and interoperability of a relational database or the type of services that the implementation can provide.

Since most microarray experiments are very complex, the design of the relational database needs to consider the possibility of creating or updating new tables. As the number of tables increase, a more complex phrasing is necessary. As this information is scattered across relations, the query process becomes dependent on the scalability of the system. Because adding and updating tables may be cumbersome, a single very large table with many attributes may be generated. However, many tuples might be empty. This is functionally equivalent to a flat file, that is non scalable.

Another main disadvantage of the R-DM is the separation of the schema from the application software. This makes updating the schema difficult. This is further complicated due to the constant evolution of biological databases and their respective schemas. To change the schema, the user needs to understand, at some level, the entire set of tables and the intricate relations of whole design. Since schemas are more valuable when they represent a clear view of the components of the database, schemas should not be affected by implementation considerations, such as limits on the number of classes, tables, or attributes. Therefore, while constructing global schemas it is necessary to detect semantic conflicts among existing tables (such as naming inconsistencies and identical entities entered multiple times).

## 3.2 Object Oriented Data Model (OO-DM)

Object-oriented programming languages are the dominant form within development environments for large-scale software systems. This is relevant in biology since many genomic projects acquire a considerable amount of data in a short period of time. Beginning in 1980's, the OO-DM was proposed to scale the access of biological and genomic information and to address some of the limitations of the relational data model [55-65]. The Object Oriented (OO) data model associate actions and functional information along with data. It has been referred to as “data with attitude.”

The OO data model *encapsulates* each *tuple* as an *object* into a single unit called *class*. Since the underlying details of a class are masked behind access methods, objects from radically different implementations can be combined in a single query. This allows the OO-DM to provide access to the data via methods or functions which can conceal a certain amount of complexity. This leads to increased portability and interoperability, since interfaces, rather than direct access to underlying data model features, are used. Since the OO-DM provides a more intuitive structure for human access, and because of its inherently modular structure, OO systems tend to be easier to maintain and reuse than purely relational ones. Also the use of object identifiers (OIDs) used to reference the accession methods in objects, makes the code more scalable. This can lead to significant performance improvements over relational databases.

Generally speaking, objects have three features: state, identity and extensibility. Identity assures that we are accessing the correct object. The state is characterized by a set of attributes (the data contained in the object) as well as any history of modification or ownership. Behavior is characterized by a set of methods that are applicable to the object. Extensibility is an especially powerful concept in software development and refers to the ability to add functionality to an existing system without fundamentally changing it. Most important is the idea that old methods of accessing the data should continue to work, even if new features are added. An object-oriented approach to programming provides extensibility in two ways: behavioral extension and inheritance. Objects may be extended by simply adding additional methods. This is tremendously valuable because developers can rely on existing behaviors in building tools that reference the information in the object. An OO approach further promotes extensibility through reuse or inheritance. It is important to note that while the terminology of the OO-DM is inspired in part by biology, the analogy is limited at best, and the biological metaphors should be taken with a grain of salt.

### 3.2.1 Notation of the OO-DM

Object-oriented notation and modeling is one of the key aspects in the development of an OO-database. During this process the use case scenarios, class/object diagrams which represent the main functionality as well as the structural aspects of the system are presented in an intuitive manner. The procedural control flow of the whole OO-database is represented schematically using standardized stereotypes (Figure 8).

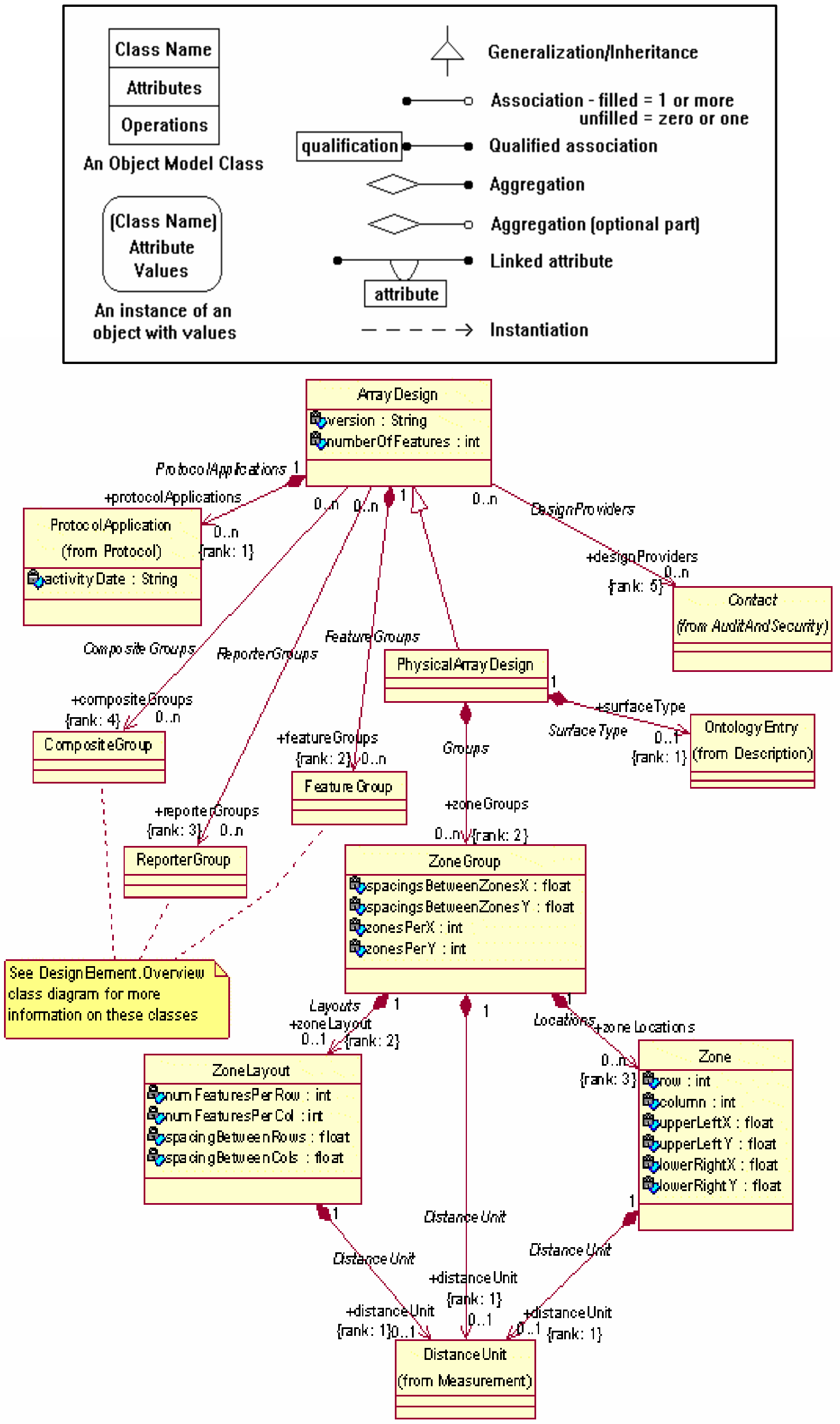

*Figure 8.* Notation of a high-level conceptual representation of the OO-RM components. The figure also includes the notation diagrams for the ArrayDesign object of the MAGE-OM

(Microarray Gene Expression - Object Model). This diagram uses the Rose Web Publisher and is implemented by the MGED and the European Bioinformatics Institute (EBI). Notice that the type of model should be indicated using the appropriate stereotype listed in top of the figure.

## 3.3 The eXtensible Markup Language (XML)

XML is derived from the Standard Generalized Markup Language (SGML), the international standard for defining descriptions of the structure and content of different types of electronic documents [60]. The XML is a data source in that its presentation is separate from its structure and content. The manipulation of genomic information using XML represents an interesting alternative and is currently implemented in different bioinformatic applications including microarray data integration efforts.

XML not only allows information in different representations to be exchanged between applications in a generic format, but also offers an opportunity to access information managed by heterogeneous DBMSs. The XML data defines the structure and content, and then a stylesheet is applied to it to define the presentation. Since XML data is stored in plain text format, XML provides a software and hardware-independent way of sharing data. Furthermore, XML can be used to represent the query results as datagrams, and XSLT (Extensible Style Language Transformation) provides a mechanism for transforming the datagrams into XML.

The relevance of the XML framework is particularly useful for re-ordering of microarray gene expression data. Since XML provides a framework for tagging structured data that can be used for specific tag sets and therefore for defining standard specifications. An XML document is either well–formed obeying the syntax of XML or XML valid conforming the logical structure defined by document type description (DTD) [60, 66]. The DTD is the classification system that defines the different types of information in any XML document. Any Web page that indicates the DTD it conforms to will instantly allow the user of an XML-enabled search engine to restrict queries to that DTD-defined space.

The Extensible Markup Language/Resource Description Format (XML/RDF) was developed by the W3C to enhance the XML model and encode metadata concerning web documents. Instead of defining a class in terms of the properties its instances may have, the RDF vocabulary describes properties in terms of the classes of resource to which they apply. XML/RDF as is, without a higher level formalism that encompasses the expressivity present in frame-based languages does not go far enough to allow the kind of modeling needed in the bioinformatics community. Three main elements are part of an XML file.

- **XML Tag:** A start tag is an element type name enclosed in angle brackets that opens an element. Every start tag must have a corresponding end tag. An end tag finishes the content of an element, comprised of an angle slash and then the element type name, all enclosed by angle brackets.

- **XML Attribute:** Attributes are name value pairs that are associated with an element type. They follow the element type name inside the start tag. They can be thought of as the 'adjectives' of XML.

- **XML Element:** An element consists of a start/end tag pair, some optional attributes defined as key/value pairs and the data between the tags.

### 3.3.1 The Microarray Gene Expression Markup Language

Microarray Gene Expression Object Management (MAGE-OM) is a data-centric Universal Modeling Language (UML) that contains 132 classes grouped into 17 packages, containing in total 123 attributes and 223 associations between classes reflecting the core requirements of MIAME [45]. MAGE-OM is a framework for describing experiments performed on all types of DNA-arrays. It is independent of the particular image analysis and data normalization algorithms, and allows representation of both raw and processed microarray data. Since MAGE-OM defines the objects of gene expression data independent of any implementation, it allows users to describe the experimental process using free-text descriptions. There are three abstract classes in MAGE-OM from which all the classes in the model derive from, Extendable, Describable, and Identifiable.

The MGED society implemented the Microarray Gene Expression Markup Language (MAGE-ML) as an XML representation of the MAGE-OM. A major advantage of the MAGE-ML format is that while it supports information from a variety of gene expression measurements including related data collection methodologies; it does not impose any particular data analysis method [45, 67, 68]. MAGE-ML also has advantages in the sense that many laboratories can verify microarray experiments with other methodologies such as real time PCR. MAGE-ML is organized into sub-vocabularies in such a way that the sub-vocabularies are independent of each other. These sub-vocabularies are driven by the packages and identifiable classes of the MAGE-OM. The MAGE software toolkit (MAGEstk) is well developed for Perl and Java applications.

### 3.3.2 Limitations of the OO-DM

OO-DMs often assume a network of computers, with processing on the back or front end, as well as intermediate tiers, caching on each level of the database. However, there are very few software systems capable to implement a full scale object oriented data model. While the OO-DM offers scalability, there are more requirements to identify accurately different classes. Therefore, the initial design is important in ensuring the future performance of the database. Without a proper management of each class, the design will not work as *per specification* and the database will be severely impaired.

## 3.4 Object-Relational Data Model (OR-DM)

Databases with an OR-DM were developed with the aim of extending the relational information with three key features of the OO-DM: inheritance, behavior and extensibility. This functionality not only permits the management of native SQL data types, but also the handling of object-oriented multimedia information (e.g. sequences, images, and video). The OR-DM is still relational because the data is stored in relations, but, loosely organized into OO hierarchical categories. As a result, the OR-DM extends the R-DM by transforming the tuple as object and the table as class. While column holds primitive data types, the class can hold data of any type of data. This allows allow attributes of tuples to have complex types, including non-atomic values such as nested relations while preserving the declarative relational access to data. This results in a very complex data structures known as LOBs (Large Objects).

Databases designed with the OR-DM are very attractive for the integration of genomic and microarray information. They are frequently used in web applications and specialized data warehouses; although a more significant impact can be seem in data federations. A database with OR-capabilities can execute complex analytical and multimedia data manipulations (i.e. images, normalized microarray data, as well sequence information), and transform these manipulations into new, complex objects make OR-DMs ideal for a research enterprise. An OR-DBMS is represented by the PIR database [69], ooTFD (object-oriented Transcription Factors Database) [59]. OR vendors provide products such Oracle, Informix, FirstSQL/J, OpenODB DB2, and Postgre Object-relational mapping.

### 3.4.1 Limitations of the OR-DM

One of the challenges in the implementation of OR-DM is the design of a modular schema capable to allow the re-use when dealing with complex structures. Moreover, the translation layer between relational and object oriented can be slow, inefficient and very costly. This can result in programs that are slower and use considerable memory.

## 4. MICRORRAY DATA STORAGE AND EXCHANGE

Once in microarray experiments are in digital format, all the components can be shared, copied, processed, indexed and transmitted from computer to computer, quickly and flexibly. The development of new technologies to store digital information are transforming the life sciences and enabling scientists to record vast quantities of data. These advances and the improvement in the sensitivity of microarray technology have motivated the development of a considerable number of specialized databases (Table 1). As the relevance of microarray experiments increases, the use of this technology for diagnostics and clinical research present a new paradigm in the storage of this information. The scientific community has been enthusiastic about microarray technology for pharmacogenomic and toxicogenomic studies in the hope of advancing personalized medicine and drug development. The US Food and Drug Administration (FDA) is proactive in promoting the use of pharmacogenomic data in drug development. This progress means that in the future, microarray data related to clinical studies and diagnostics need to comply with regulations mandating data preservation and access.

The scope of different databases provide user with a variety of services and maintaining specific types of information associated with microarray experiments. These databases can store at least five levels of information: 1) the scanned images (raw data) 2) quantitative outputs from image analysis 3) normalized data or 4) a list of important genes after the analysis process and 5) the metadata associated with each experiment.

Microarray raw data (images) are the starting point of the analysis process. However, storing this information poses practical limitations including the size of and access to the image files. Nonetheless, considering the ongoing development in image analysis software, the storage of any processed form of the original image, without keeping the original image itself, can lead to the argument that the data is outdated as new image analysis methods become available. In early 2001, there was considerable discussion about who should maintain original microarray images and if this was responsibility of journals, public repositories, or research institutes. Despite the intense debate, no consensus has been reach about whether or not is cost-effective to store all this information, and, at this point, the publishing authors themselves are responsible for storing (and providing on request) original image files. Certainly, no decision has been made regarding if this task should be ensured by public repositories or the institutions hosting the author of a particular paper [35, 67].

Sharing the extracted (but not normalized i.e. CEL, GPR files) solves some of the practical limitations related with raw images. This level of data level sharing is well suited for many microarray public and local databases. However, it requires the implementation of appropriate DBMS as well pre-processing tools. Another approach to store microarray data consists in the sharing of normalized expression ratios or summarized values such as signal intensities. In this form, much information about the experiment is lost because the diversity of microarray data normalization and probe level analysis techniques. The last form of microarray data exchange consists in providing a list of genes that significantly differ between experimental samples. Due to the wide variability in accuracy across different analysis methods, this information should be limited only to publications. Finally, the sharing of microarray metadata is another component of the data exchange process; however, it has not received considerable attention. Supplying metadata to describe microarray experimental details is not a rewarding task, since it requires considerable work and there is not an immediate benefit. Considering that microarray experiments are done by different communities and have different scope, we can classify these implementations as:

- **Public**: These types of microarray DBMSs cover different microarray experiments by single or different researchers, and allow users to query, retrieve and analyze both unpublished and published microarray information.

- **Institutional**: The configuration of this type of microarray DBMS resembles public databases but is built around a particular organism and/or restricts the access to a limited number of researchers depending on some set of permissions defined by the institution.

- **Private**: These microarray DBMSs are limited to researchers within a research group and are not available to other researchers.

*Table 1.* Microarray Data Storage and Exchange Systems

| Database Name | Schema | Public | Local | MIAME Supportive | Software |
|---|---|---|---|---|---|
| Acuity | No | No | Yes | Yes | SQLServer |
| AMAD | No | No | Yes | No | Flat File, PERL |
| AMANDA | No | Yes | Yes | Yes | MySQL |
| Argus | No | No | Yes | No | MS Web-server |
| ArrayDB | Yes | Yes | Yes | Yes | Sybase, PERL |
| ArrayExpress | Yes | Yes | Yes | Yes | MySQL, Oracle |
| Axeldb | Yes | Yes | Yes | Yes | Perl, FlatFiles |
| BASE | Yes | No | Yes | Yes | MySQL, PHP |
| BioMart | Yes | Yes | Yes | Yes | SQL |
| CGO | No | No | Yes | No | MS-Access |
| CoBi | No | No | Yes | Yes | Oracle |
| Dragon | Yes | Yes | Yes | No | MS-Access |
| ExpressBD | Yes | Yes | Yes | Yes | Sybase |
| Expressionist | No | No | Yes | Yes | Oracle8i, Web server |
| GeneDirector | No | No | Yes | Yes | Oracle |
| Genetraffic | No | No | Yes | Yes | PostgreSQL |
| GeneX | Yes | Yes | Yes | Yes | PostgreSQL, XML |
| GEO | Yes | Yes | No | Yes | Flat files |
| GeWare | Yes | Yes | Yes | Yes | Flat File, OLAP |
| GXD_GEN | Yes | No | No | Yes | Flat File |
| LIMaS | Yes | No | Yes | Yes | Flat File, Java |
| MADAM | Yes | No | Yes | Yes | Flat File, XML |
| mAdb | Yes | No | No | Yes | Sybase |
| MARS | Yes | No | Yes | Yes | MySQL, Oracle 9i |
| maxdSQL | Yes | No | Yes | Yes | MySQL, Oracle8i, |
| M-Chips | Yes | Yes | Yes | No | PostgreSQL |
| MEM-5AM | No | No | Yes | Yes | DB2 |
| MEPD | No | Yes | No | No | DB2 |
| NASCArrays | Yes | Yes | No | Yes | FlatFile, XML |
| NOMAD | No | No | Yes | No | MySQL |
| OrionDB | No | No | Yes | Yes | PostgreSQL, XML |
| PartisanLIMS | No | No | Yes | Yes | --- |
| RAD | Yes | Yes | Yes | Yes | Oracle, PHP |
| READ | No | No | Yes | No | FlatFile, PostgreSQL |
| RossettaResolver | No | No | Yes | Yes | Oracle |
| SGMD | No | No | Yes | No | SQLServer2000 |
| SMD | Yes | Yes | No | Yes | Oracle, PostgreSQL |
| SMD | Yes | Yes | Yes | Yes | Oracle, Perl |
| StressDB | No | No | Yes | Yes | Oracle |
| Longhorn Array | Yes | Yes | Yes | Yes | PostgreSQL |
| YMGV | Yes | No | Yes | No | PostgreSQL, PHP |

## 4.1 Microarray Repository

Microarray data repositories are data collections that, in general, are implemented by one institution to serve a research community [61]. These storage and exchange systems allow the submission of data from both internal and external investigators [70, 71]. Although often used synonymously with data warehouse, a repository does not have the analysis functionality of a warehouse. The maintenance and curation of data repositories has made these data exchange systems of considerable value to specific research communities. Since repositories need to be able to store, access, filter, update and manipulate large data sets quickly and accurately, the information requires systematic knowledge management, proper representation, integration and exchange.

## 4.2 Microarray Data Warehouses and Datamarts

Data warehouses are databases devoted to storing relevant information from other sources into a single accessible format [72]. These systems have the advantage that they can import and analyze data that cannot otherwise communicate with each other. Since they incorporate a time factor, data warehouses can present a coherent picture of heterogeneous genomic sources integrated at different time points. In fact, very often, their requirement is to capture the incrementally changed data (delta) from the source system with respect to the previous extract.

Data warehouses are populated from the primary data stores in three main steps often through sophisticated compression and hashing techniques. First, data are extracted from the primary data sources. This process uses monitors/wrappers that are capable of both collecting the data of interest and send it to the warehouse. The monitor is also responsible for identifying changes in external databases and updating the warehouse automatically. Second, the data are transformed and cleaned. Specific logic for data standardization or for resolving discrepancies between data can be implemented in this step. Third, the data are loaded into the database, and indexes are built to achieve optimum query performance. This configuration facilitates the direct access of microarray data for analysis, allowing for both good performance and extensive analysis and visualization capabilities.

In order to standardize data analysis, data warehouses are organized as problem-driven small units called datamarts. These implementations are subsets of larger data warehouses and contain data that has further been summarized or derived from a main data warehouse. Datamarts are an attractive option because they take less time to implement than a centralized data warehouse and initially cost less. However, data marts can be more costly in the long run because they require duplicated development and maintenance efforts, as well as duplicated software and hardware infrastructure. Additionally, scattered data marts can hinder enterprise performance because they often store inconsistent data, making one version of "the truth" impossible to obtain.

The development of data warehouses like MaxD and DataFoundry which integrated SwissProt, PDB, Scop, Chat and dbEST in a unified schema represent a clear success of genomic data warehousing [72, 73]. First, it must obviate the need for the conversion and migration of data and must require no change to the local database system. Second, it must allow users to interact in such a way that both users and applications are shielded from the database heterogeneity. Third, allowing the interoperability of heterogeneous databases must allow reads and updates of these databases without introducing changes to them. By their nature, data federations (datamarts) do not modify the primary data sources and a great effort must be paid in the cleaning and transformation before their placement in the warehouse. Since data are drawn directly from the primary data stores, detection and cleaning of redundant data is not easily incorporated [74, 75].

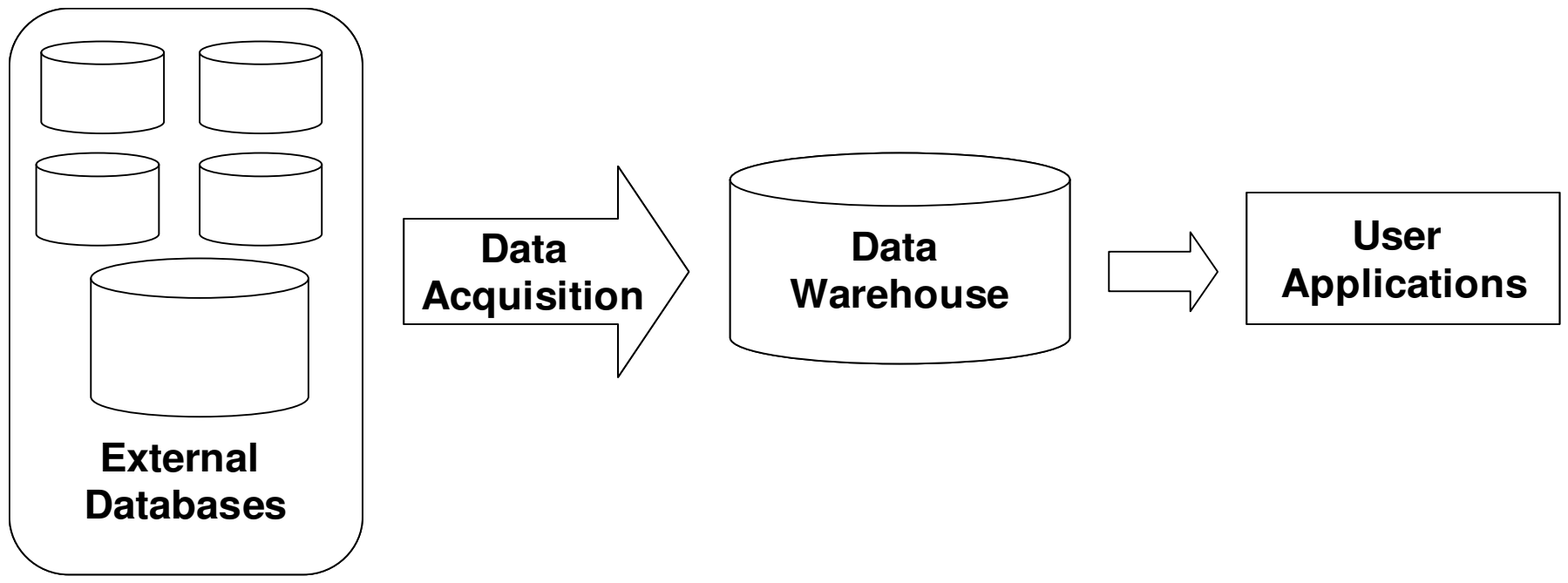

*Figure 9.* Data warehousing approach: Several data extraction components move and integrate the data into the virtual warehouse. Different software applications can be used to analyze and represent microarray information and integrate it with other genomic data sets.

Microarray data warehouses have three two costly drawbacks: 1) considerable effort required for planning the integration and 2) the great deal of investment required for data cleaning and transformation. This situation affects reliability and overall system maintenance of the system.

## 4.3 Microarray Data Federations

Most microarray databases are specialized collections of information for a particular organism or biological process. They are scattered in different locations and managed under difference policies. In order to integrate this information, a data federation schema seeks to join isolated, heterogeneous repositories into a single virtual main database. This process is accomplished without modifying the primary data sources and by avoiding the creation of a large warehouse. Their use is motivating the emergence of "virtual organizations" which take advantage of the standardization of microarray protocols and the use of reference probes. In addition, federations rely on the development of GO, MO, MIAME and MAGE-ML standards which permits the development of wrappers exploring and querying multiple data sources and may have different characteristics including:

- **Public data:** data from public sources, such as ArrayExpress and NCBI-GEO, copies of raw data may be held locally for performance reasons or shared throughout the federation.
- **Processed public data:** public data that has additional annotation or indexing to support the analyses needed by different analysis algorithms. This information can serve as the common link for joining different databases within the federation.
- **Sensitive data:** In many cases, an individual user will be generating data which remains to be analyzed or is unpublished. These require careful enforcement of privacy and may be restricted to one site, or even part of a site.
- **Personal research data:** data specific to a researcher, as a result of experiments or analyses that that researcher is performing. This is not shared even among the local team. It may later become team research data.
- **Team research data:** data that is shared by the team members at a site or within a group at a site. It may later become consortium research data, e.g. when the researchers are confident of its value or have written about its creation and implications.
- **Consortium research data:** data produced by one site or a combination of sites that is now available for the whole consortium.

While data federations could accelerate the development o data standards, traditional federations might be too rigid and labor-intensive to adapt to an open environment where new sources are integrated dynamically. Therefore, before implementing or joining a data federation, researchers interested in this possibility need to address issues related with the design, interoperability and security of each transaction and the transfer of high volumes of information.

In most cases, member databases are geographically distributed, hosted on a variety of platforms and administered independently according to differing policies, which might be independent of the federation policies. This means that the federated system must be designed under the assumption that not all resources will be available and consistent at all the times. This makes the quality control very difficult. Because data federations perform a considerable number of data transformations, query performance is one of the main concerns.

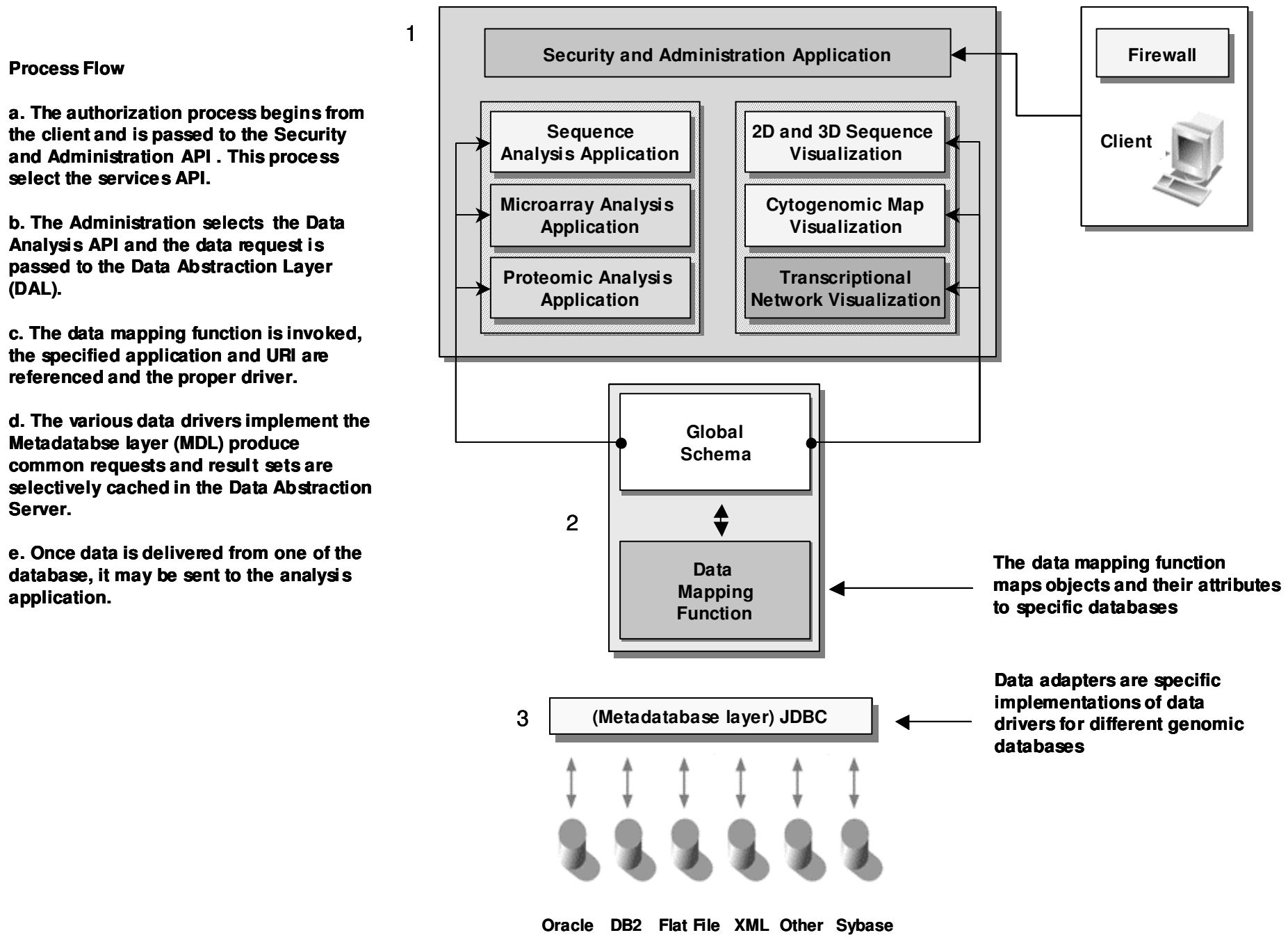

*Figure 10.* Basic schematic representation of an enterprise federated system for microarray and other genomic information integration and management: 1) application layer, 2) abstraction layer and 3) metadatabase. Notice that the figure illustrates the various components and services based on a J2EE environment. This allows the user to make complex queries across multiple internal and external biological data sources.

## 4.4 Enterprise microarray databases and M-KM

From an institutional perspective, the information generated by microarray experiments can be interpreted as data, values, and relations generated by different researchers with a shared common and main institutional goal [76, 77]. In this context, enterprise systems can be defined as computer architectures designed as intranet systems capable of performing pipeline operations. Using specific hardware, software, database management systems, agent software, analysis and visualization algorithms enterprise systems integrate information and find patterns and relations over large periods of time. The result of this process is the constant transformation of data into an intellectual asset (Figure 11). The implementation of enterprise microarray data management systems is being enhanced by the development of semantic web technologies, grid computing and Internet2 the capacities.

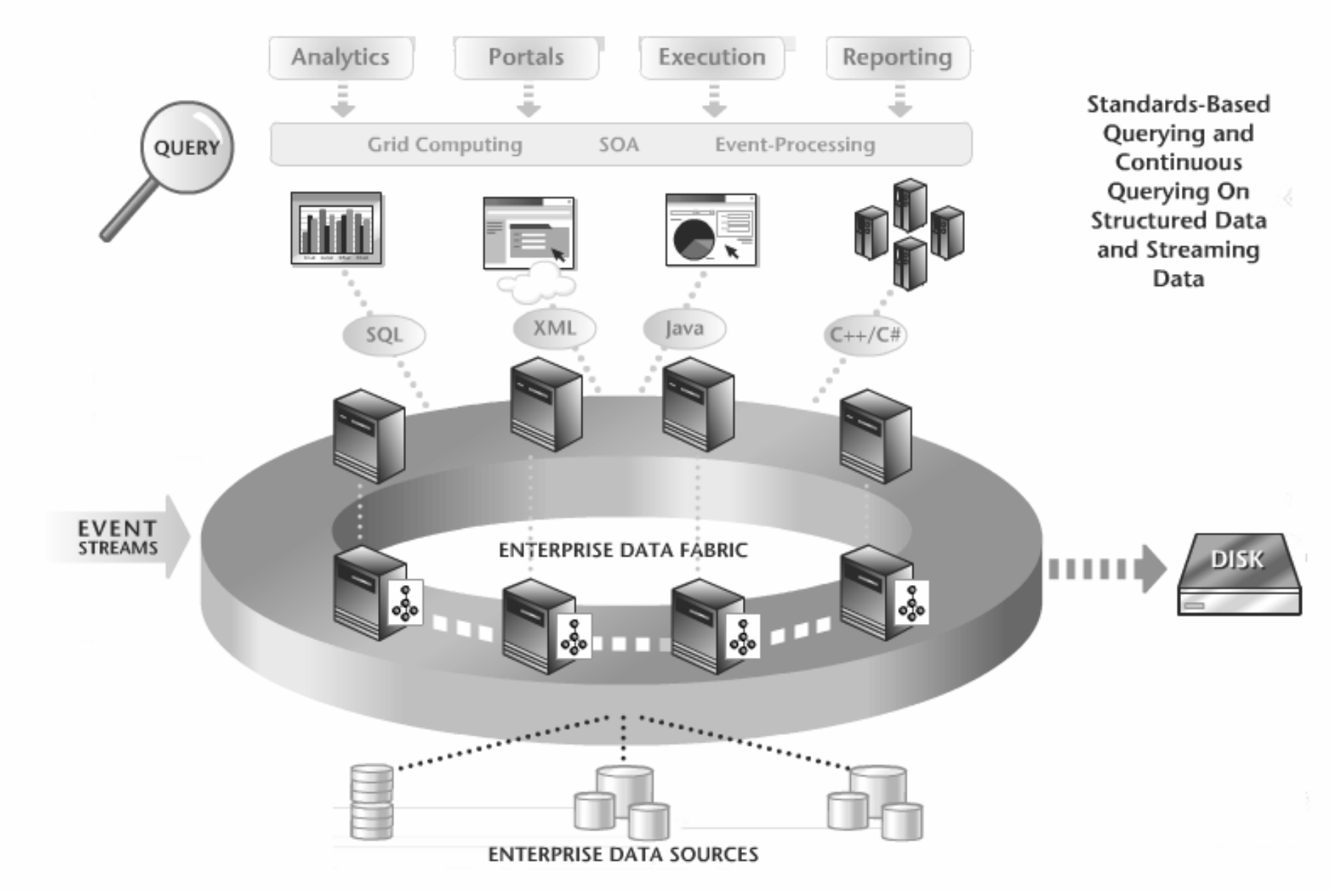

*Figure 11.* Enterprise Knowledge Management System for Microarray Data.

The implementation of enterprise microarray data management systems is resulting in a new generation of infrastructures known as knowledge management systems (KMS). The KM concept evolved from information management tools and not only to integrate data, but integrates many aspects of computer-supported collaborative work environments including blogs, wikies and discussion forums. Difference from conventional databases, microarray KM tries to consolidate knowledge which is not easily codified in digital form, such as the intuition of key individuals with considerable experience interpreting data from a particular biological process, organism or cellular process. These individuals an/or their collective thinking might recognize various patterns of gene expression profiles that someone with less experience or a single individual may not recognize. While promising, microarray KM implementation requires a series of standards to enable genomic information to be captured, analyzed, understood and re-applied in new contexts. This includes detailed technical and microarray metadata, learning management tools, content modularization, genomic data analysis workflows, supervised and unsupervised analysis algorithm and visualization. Therefore, the implementation of an enterprise analysis system requires:

- **Technical integration:** Use nonproprietary platforms, open standards and methodologies in the design of the system architecture that ensure long-term scalability, robustness, performance, extensibility and interoperability with other systems and platforms.

- **Semantic integration:** Use all levels of linked biological concepts and their dependencies in biological, genetic and microarray ontologies. Manual intervention to map data between different data sources should not be required.

- **Interoperability:** Provide user with the ability to directly import and export gene expression data as a single flat files derived from separate microarray DBMSs.

- **Allow configurable combinations of data sources:** It should be possible to integrate and combine different sources of biological information.

- **Evidence management:** It should be possible to determine which computational method was used for derived data and to annotate mappings between databases.

Microarray KM systems can be technically similar to data federations, however, their unique features make these systems ideal for reinforcing intra and multi-organizational data sharing, and to validating the description of the molecular organization and dynamics of specific biological processes. Since this information becomes takes advantage of the collective knowledge available from researchers within a particular research enterprise, the goals and planning of the organization can be optimized (Figure 12). The collaborative foundation of microarray KM is beginning to attract small research groups to integrate their applications into easily and seamlessly accessible platforms that can be used in a open environment.

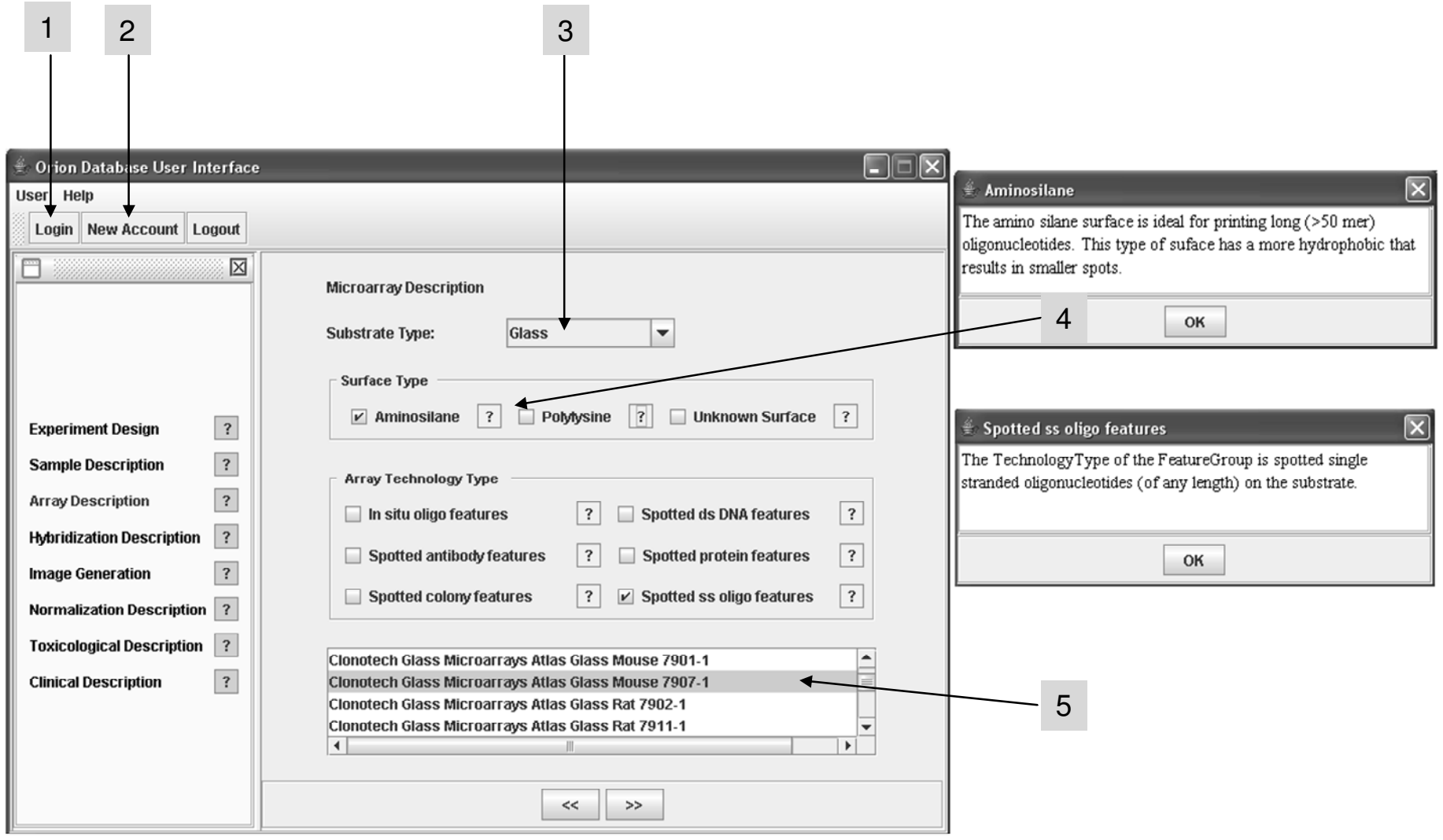

*Figure 112.* Graphical user interface (GUI) of OrionDB. This solution is implemented by Orion Integrated Biosciences Inc. as a microarray data storage and management system for research enterprises. When the user is logged into the system (1), the system assigns a personalized GUI (2) and provides access to specific projects associated with the user's privilege profile. When the user is inserting data, automatic data structures are generated (3 and 5). In case the user needs to clarify a particular concept to annotate microarray data, MO based concepts can be generated by clicking on the integration symbols (4). Each microarray platform (5) stores metadata and data associated with each spot. This information is mapped to genomic and other molecular biology databases.

## 5. CHALLENGES AND CONSIDERATIONS

Microarray technology has added an important dimension and depth to the analysis of different and dynamic different biological processes. The scientific value of this technology is enormous; however, the quality of this information is highly variable. Problems in data quality have been observed from analyzing published data sets, and many laboratories have been struggling with technical troubleshooting rather than generating reliable datasets. Therefore, it is important to recognize that not all datasets are suitable for storage and distribution. Unless a clear description of the experimental design and quality experiment itself is provided (technical and biological replicates, and the use of appropriate protocols) the query and retrieval of datasets should be limited to published results. The fact that many of these datasets do not provide appropriate metadata makes difficult the incorporation of quality assessment methods. Therefore, it is necessary to implement semi-automated approaches that score the level of reliability of the data. Developing better systems for collecting metadata, either manually or automatically is one of the most urgent issues needing attention.

Several microarray databases and analysis software overcome national boundaries. This is particularly true in the sharing of microarray data, where scientists on a global basis deposit and retrieve data irrespective of who funded the information production. Some microarray databases have already surpassed a terabyte scale. The implications of the accumulation of this information, has been not fully recognized. There are several critical design issues in databases which affect how new databases and analysis systems are implemented. Performance and efficiency not only needs to be measured by query response time, but by the time it takes a scientist extracts knowledge from the data. Adopting standards which are likely to survive and/or are well described for the future is difficult. Therefore, it is necessary to motivate the re-use of software and the development of approaches to decrease the risk of data loss or the expense of data resurrection.

Large data repositories, computationally intensive data analysis and visualization tools pose difficult problems for the implementation of open access enterprise microarray data management and KM systems. Commonly, database schemas are changed without any notification, explanatory documentation, or appropriate notation. This makes the maintenance and improvement of these systems difficult. These challenges are complicated by the fact that internet bandwidth and data compression technologies have not kept pace with the growth of scientific data sets. Many data repositories still provide data access primarily via FTP. While FTP-based data sharing is a valuable starting point, we need to encourage more robust interfaces, capable of retrieving specific datasets automatically. This is perhaps a main bottleneck in the automatic retrieval of databases since there is poor communication on the part of the resource maintainers. Moreover, large data archives are becoming increasingly 'isolated' in the network sense. Therefore, in order to work with large data sets, it might be necessary to send computations to the data, rather than copying or moving the data across the internet.

A limiting aspect in the development of microarray data storage and exchange systems is related to the complexity and dynamics of the data itself. Complexity arises from the lack of unique spot identifiers and the existence of a large number of many-to-many relationships among clones, accession numbers, chromosomal location, mutation types, etc. In addition, microarray datasets derive the treatment of biological samples (with different genetic background) to multiple experimental conditions and time courses. The dynamics of microarray data results from the terminology used for the description of a biological sample and the functional role for a particular gene or its transcriptional variants. These attributes can change as new discoveries update this information. As a result, the interpretation of a particular microarray dataset is highly dependent on ever-growing and dynamic annotation information. Although the use of microarray data analysis tools is beyond the scope of this chapter, the reader should be aware that the annotation of unknown genes using ontologies depends on analysis algorithms and the amount of information used in the analysis process. It is now more evident that the "guilt by association" is not always true.

The reader must be aware that deciding on appropriate terms that could be used in the development of microarray ontologies and mapping them to other middle and upper ontologies entails main decision points. First, the implementation of a large and comprehensive ontology versus several smaller task oriented ontologies is still a subject of discussion. One alternative (large ontologies) presents challenges regarding agreement across sub-disciplines. Second, coordination between different small ontologies could be very expensive. In both situations, it is necessary to consider how the dynamics of the ontology will affect a database. This is important in biological ontologies because they do not remain static; they evolve as new discoveries are made. By restricting access to or simplifying assumptions about a particular dataset in order to accommodate it to a particular ontological definition, the user risks the trivializating the queries and results.

The reader should be cautious during the integration of different microarray datasets and the annotation of new genes based on combined gene expression values. Simplistic, linear transfer of derived information can lead to a "transitive catastrophe" or "data poisoning," in which one piece of inaccurate information can corrupt a large number of derived results. This legacy issue is becoming more significant since the functional inference of genes and transcriptional interactions changes with time and is not straightforward. As more microarray data becomes available, it is becoming evident that biological systems are organized as transcriptional networks with specific modular components, rather than in a particular class or cluster of similar gene expression values.

## 6. CONCLUSIONS

Since the early 1990s, when scientists first began using microarray devices to study gene expression, they have widened the use of this technology to studying how genes interact at the transcriptional, proteomic and metabolomic levels. The rapid increase in the size and diversity of this type of information has highlighted the need for efficient computational techniques for data storage and exchange. The internet has made it possible to access large amounts of information from multiple microarray databases distributed across the world. This is stimulating a growing demand for analysis and visualization systems of multiple and heterogeneous biological data sources. However, even when a global network infrastructure provides the foundation for the microarray data sharing and exchange, the location, retrieval and the combination of disparate and poorly annotated microarray datasets has proven to be a complex and a time consuming task.

Researchers recognize the benefits of integrating microarray with other genomic information. Investing in these efforts, are not only saving time, but also making more effective experimental designs and reducing experimental resource expenses. Due to the large number of data points and since the analysis of the same data using different computational techniques can lead to a better understanding of the biological process, different microarray data repositories are playing a vital role in biological sciences. *Data exploration research* is now impacting traditional wet-lab experiments from hypothesis generation to experimental design and data analysis. However, how good genomic data mining is made dependents on the time and care that is spend when designing and implementing a data storage and exchange system. Specially now, that a new generation of researchers no longer 'do' wet-lab experiments. Instead they 'mine' available microarray databases, looking for new patterns and discoveries.

The integration of data is an active research field in the computational sciences. However, as new technologies collect large amounts of genomic information in a near real time fashion, the storage and exchange of data streams will continue to challenge a new generation of researchers. Therefore, important questions in database design will need to be addressed. The inclusion of different data types and the communication with other very large databases will be one of the most important challenges for an integrated initiative toward the understanding of complex biological systems.

# 7. ACKNOWLEDGMENTS

The authors would like to thank Dr. Eric M Blalock at the University of Kentucky - Department of Molecular and Biomedical Pharmacology for providing insightful suggestions and editorial comments.